\documentclass[aps,pra,showpacs,twocolumn,groupedaddress]{revtex4-1}

\usepackage{graphicx,amsmath,amssymb,mathtools,xcolor}
\numberwithin{equation}{section}

\definecolor{deepred}{HTML}{CD2626}

\begin{document}
\title{Betweenness Centrality in Dense Random Geometric Networks}
\author{Alexander P. Giles}
\email[]{Alexander.Giles@bristol.ac.uk}
\address{School of Mathematics, University of Bristol, University Walk, Bristol BS8 1TW, United Kingdom}
\author{Orestis Georgiou}
\email[]{Orestis.Georgiou@toshiba-trel.com}
\address{Toshiba Telecommunications Research Laboratory, 32 Queen Square, Bristol, BS1 4ND, United Kingdom}
\author{Carl P. Dettmann}
\email[]{Carl.Dettmann@bristol.ac.uk}
\address{School of Mathematics, University of Bristol, University Walk, Bristol BS8 1TW, United Kingdom}

\date{\today}

\begin{abstract}
 Random geometric networks consist of 1) a set of nodes embedded randomly
in a bounded domain $\mathcal{V} \subseteq \mathbb{R}^d$ and 2) links formed probabilistically according to a function of mutual Euclidean separation. We quantify how
often all paths in the network characterisable as topologically `shortest' contain a given node (betweenness centrality), deriving an expression in terms of a known integral
whenever 1) the network boundary is the perimeter of a disk and 2) the network is extremely dense. Our method shows how similar formulas can be obtained for
any convex geometry. Numerical corroboration is provided, as well as a discussion of our formula's potential use for cluster head election and boundary detection in densely deployed wireless ad hoc networks.
\end{abstract}

\maketitle

\section{Introduction}

Betweenness centrality $\gamma(\kappa)$ is a graph theoretic measure of
how often a node $\kappa$ is on a shortest path of links between
any pair of nodes \cite{freeman1977}. Ubiquitously
\begin{equation}\label{e:1}
\gamma(\kappa)=\frac{1}{2}\sum_{i}\sum_{j}\frac{\sigma_{ij}(\kappa)}{\sigma_{ij}}
\end{equation}
where the sum requires $i\neq j\neq\kappa$: $\sigma_{ij}$ is the
total number of shortest paths that join $i$ and $j$ and $\sigma_{ij}(\kappa)$
gives the number of those geodesics that pass through $\kappa$. Intuitively, nodes with high betweenness can be thought of as decisive for the functionality of decentralized communication networks, since they typically route more data packets (based on the assumption that traffic tries to follow only the shortest available multi-hop paths). This notion of importance is in sharp contrast to traditional methods, which simply enumerate node degrees: a bridging node which connects two large clusters is, for example, of crucial importance to the whole network, even though it may only have two neighbours; this sort of information is brought out by $\gamma$, but usually goes undetected.

In router-based communication networks, the router itself has a normalised betweenness of unity, since all nodes connect to it directly, while all other nodes have a centrality of zero. A promising focus in physical layer network design today is, however, on an entirely different network philosophy, where there is no router \cite{dettmann2007,santi2003,li2009}. These structures are known as wireless ad hoc (or sometimes `relay') networks, where packets of information are routed in a multi-hop fashion between any two nodes that wish to communicate, allowing much larger, more flexible networks (due to the lack of pre-established infrastructure or the need to be within range of a switch). Commercial \textit{ad hoc} networks are nowadays realised under Wi-Fi Direct standards, enabling device-to-device (D2D) offloading in LTE cellular networks \cite{asadi2015}.

This new diversity in machine betweenness can be harnessed in at least three separate ways: historically, in 2005 Gupta \textit{et al.} \cite{gupta2005} used $\gamma$ as a criteria for electing cluster head nodes which communicate to base-stations on behalf of all the cooperating machines, and later, in 2010 Ercsey-Ravasz \textit{et al.} \cite{ercsey2010} demonstrated how betweenness can be used to delineate the `vulnerability backbone' of a network (a percolating cluster of the highest $\gamma$ nodes), which is important for defense purposes \cite{holme2002,dallasta2006}. Finally, in 2006, Wang \textit{et al.} \cite{wang2006} researched the use of betweenness for boundary detection (since at high node density $\rho$ the betweenness of machines exhibits a bi-modal behaviour and can therefore elucidate boundary location). Since the principal model for \textit{ad hoc} networks has become the random geometric graph \cite{gilbert1961,penrosebook} (consisting of a set of nodes placed randomly in some domain $\mathcal{{V}}\subseteq\mathbb{R}^{d}$, mutually coupled using a connection law based on their Euclidean separation), in this paper we begin to develop an understanding of how the expected betweenness of a node at some domain location changes with the parameters of the random graph model, evaluating analytic formulas for $\gamma$ as a function of domain position.

We start our derivation with the disk domain $\mathcal{{D}}$ of radius $R$ (left panel, Fig. 1), considering the limiting scenario of infinite node density with a vanishing node-to-node connection range. We will then argue that betweenness, a computationally heavy operation with possibly high communication overheads, can be well approximated by our analytical closed form predictions and can therefore prove useful in practice.

This paper is structured as follows: in Section II we present our basic network model and state our main assumptions. In Section III we introduce an analytic formula for $\mathbb{E}(\gamma\left(\epsilon\right))$ in the continuum limit (where the node density  $\rho\rightarrow\infty$), which is our main result. In Section IV we present Monte Carlo simulations which validate our predictions, in Section V we discuss the applicability of the derived betweenness centrality formula within \textit{ad hoc} wireless networks and conclude in section VI, discussing the impact of our contribution and possible future research directions.

\begin{figure}[!t]
\hspace{5mm}
\includegraphics[scale=0.39]{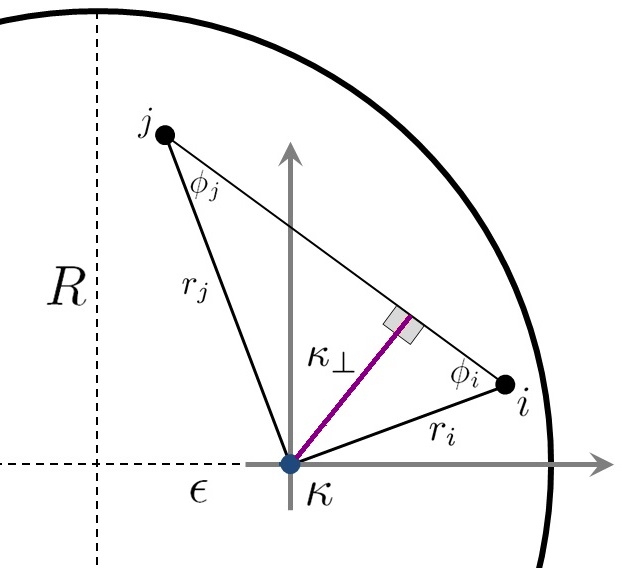}
\caption{Defining $\kappa_{\bot}$: If we
consider the three general positions $\mathbf{r_{\mathrm{i}}}$, $\mathbf{r_{\mathrm{j}}}$
and $\mathbf{r}_{\mathrm{\kappa}}$ (corresponding to the positions
of the respective nodes $i$, $j$ and $\kappa$), we have the scalar
$\kappa_{\bot}$ representing the distance of $\kappa$ to the line joining $i$ and $j$. The axis are
centred on the node $\kappa$, while the circle is centred at $\left(-\epsilon,0\right)$.}
\label{fig:one}
\end{figure}

\begin{figure*}[!t]\label{fig:1}
\hspace{5mm}
\includegraphics[scale=0.2751]{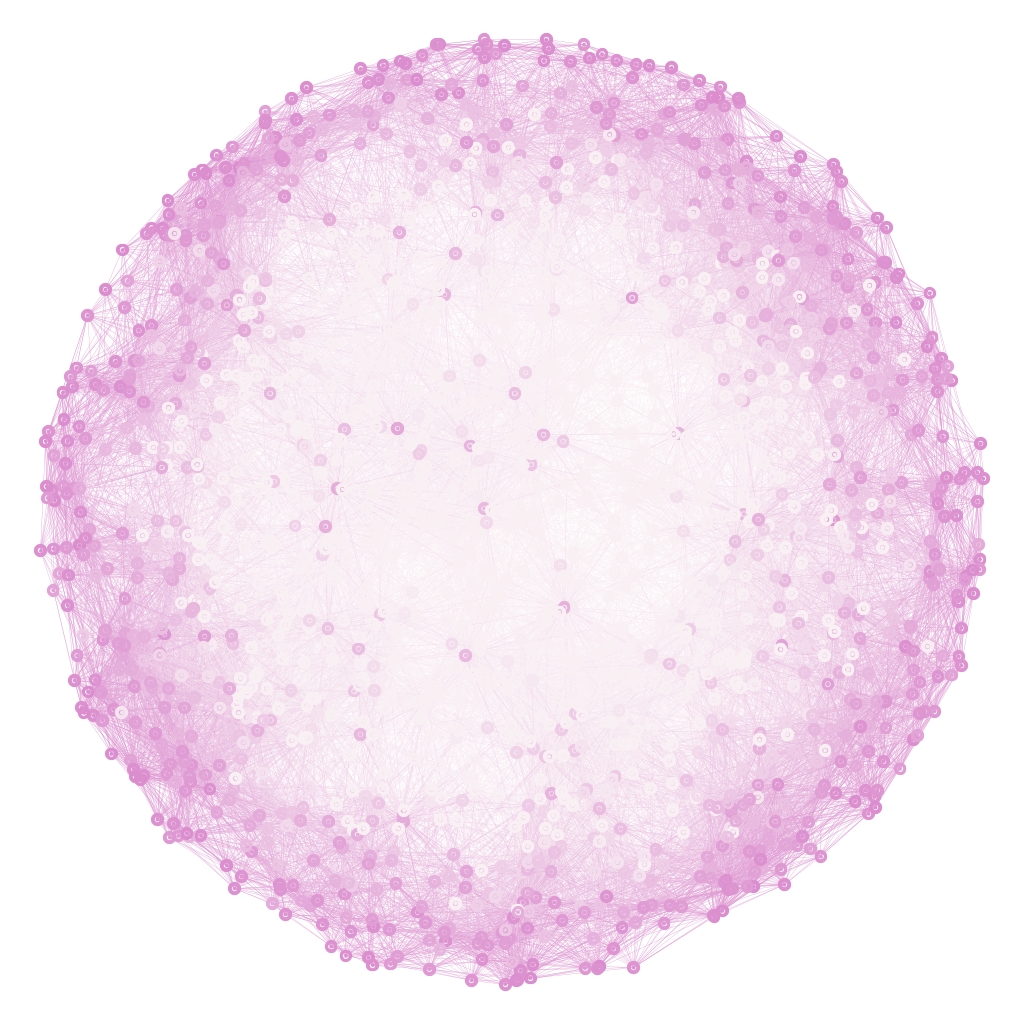}\qquad\qquad\qquad
\includegraphics[scale=0.2751]{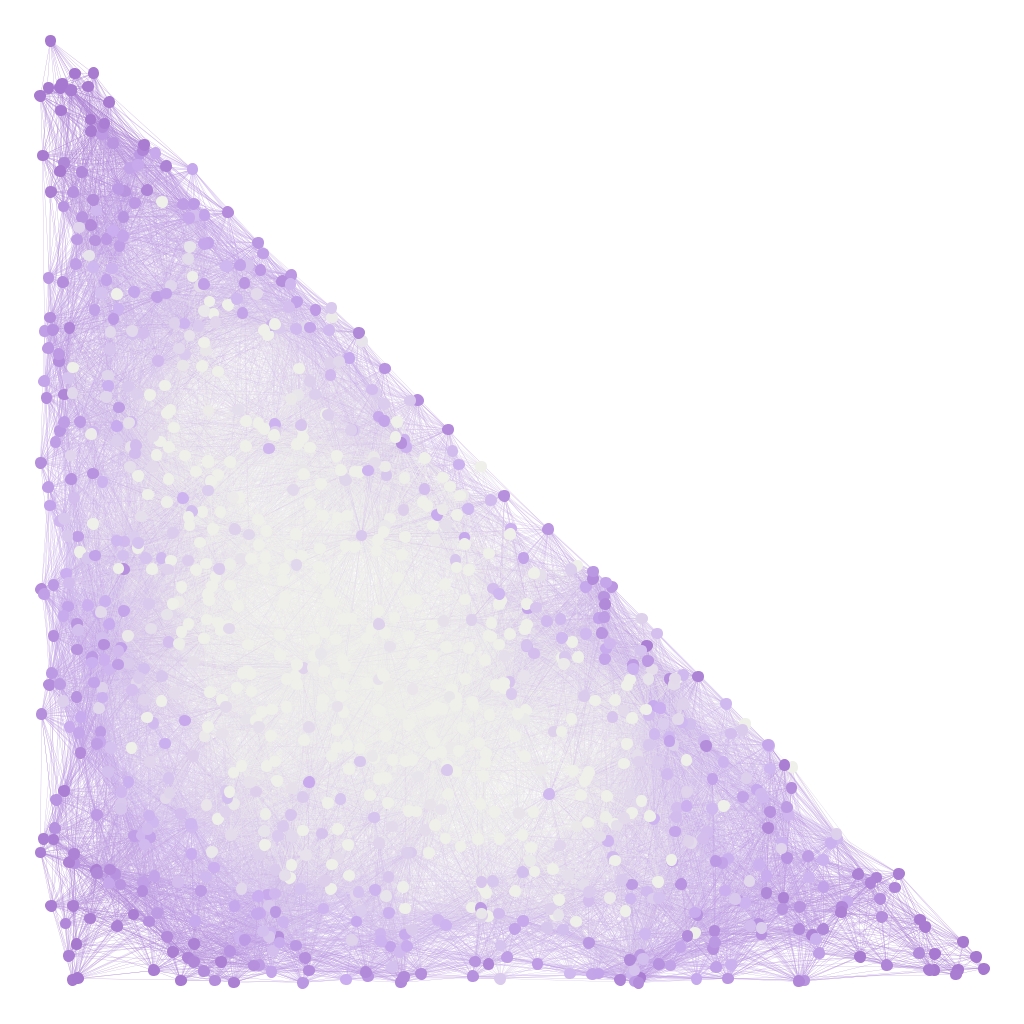}
\vspace{5mm}
\end{figure*}
\begin{figure*}[!t]
\hspace{5mm}
\includegraphics[scale=0.2751]{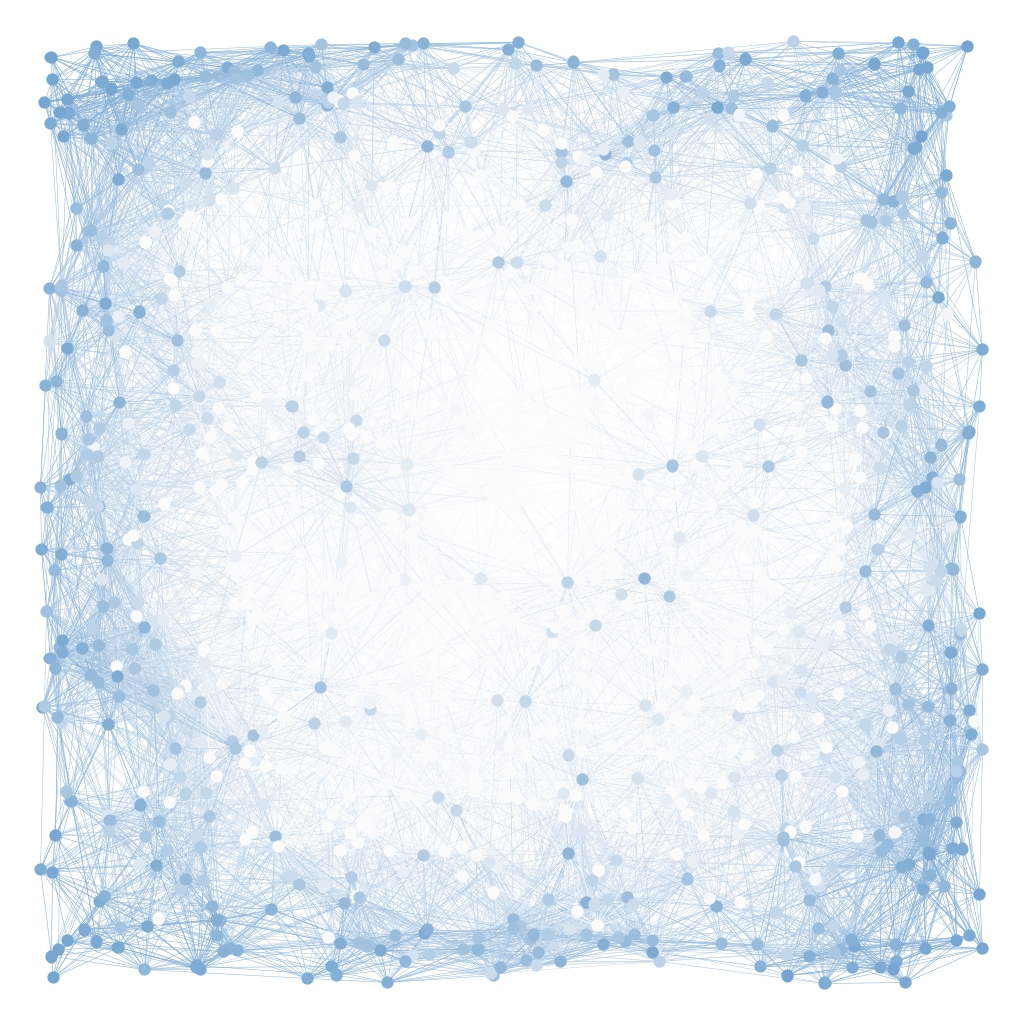}\qquad\qquad\qquad
\includegraphics[scale=0.2751]{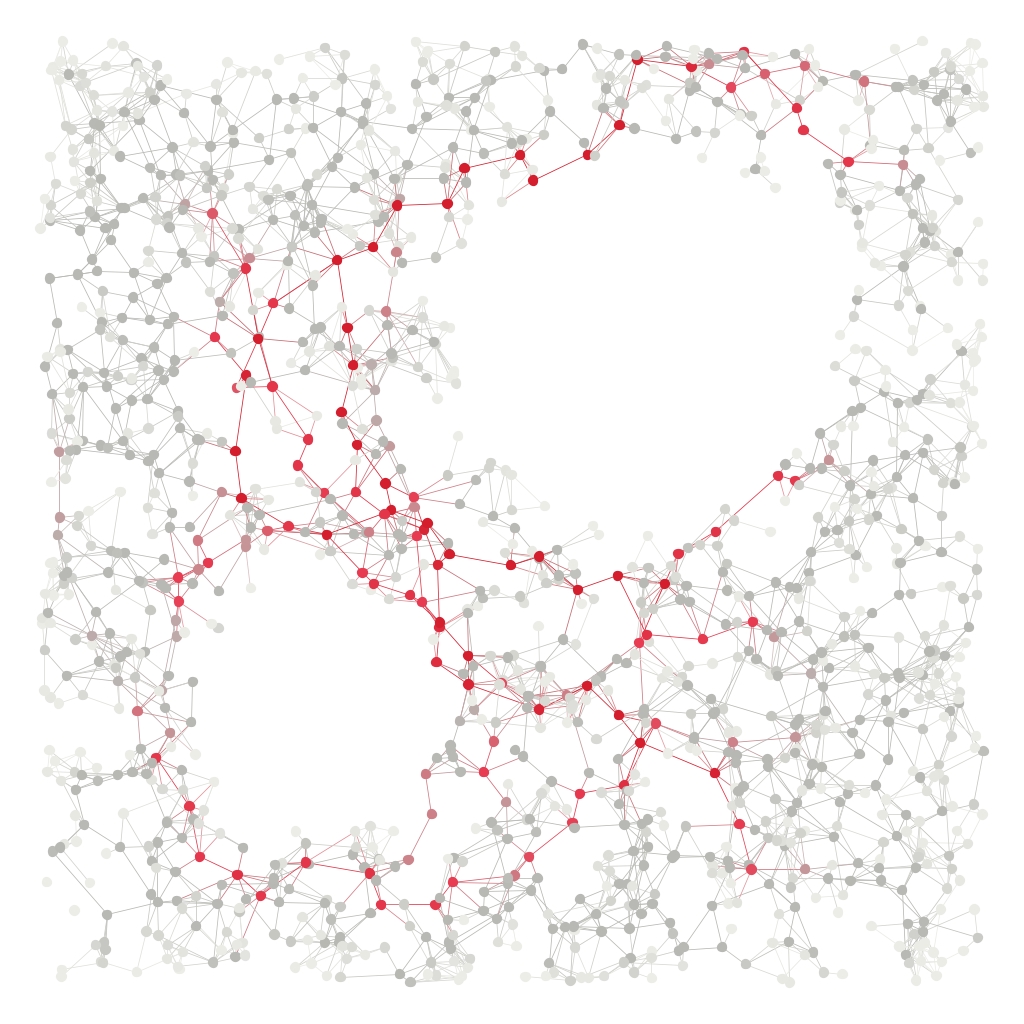}\newline
\caption{(Colour Online) Four realisations of soft random geometric graphs and their betweenness centrality bounded within various domains, including the disk $\mathcal{{D}}$, square, right-angled triangle and square domain containing two circular obstacles: in both the left and upper right figures the darker colour represents low centrality, whereas the lighter colour high centrality, whereas in the obstructed square domain (lower right) the least central nodes are faded to grey and the most central are highlighted in red. Note that the boundaries of the domains are locations where betweenness is at a minimum. The link colours are based on the average betweenness of the two connected nodes.}
\label{fig:two}
\end{figure*}

\section{Our Model}

Consider $N$ nodes placed inside a bounded, convex subset $\mathcal{{V}}\subseteq\mathbb{R}^{d}$ of volume $V$ (using the Lebesgue measure) according to a uniform point process of density $\rho=N/V$ at positions $\mathbf{r}_{i}$, $i\in\left\{ 1\ldots N\right\}$. Nodes $i$ and $j$ (at $\mathbf{r}_{i}$ and $\mathbf{r}_{j}$) possess Euclidean separation $r_{ij}$ and are connected (through a `link') with probability $H(r_{ij})=e^{-\beta r_{ij}^{\eta}}$ (where $\beta$ is a constant determining the typical node-to-node connection range \cite{cef2012}). This connection function helps to model the fact that over a wireless channel with Rayleigh fading \cite{tsebook}, the complement of the information outage probability between nodes $i$ and $j$ decays exponentially with the distance $r_{ij}$ raised to some power, the path loss exponent, which we set here equal to 2 since we consider only free-space propagation \cite{cef2012}. The resulting random graph is called `soft' due to the probabilistic connection law \cite{penrose2013}, a generalisation of the more common `hard' unit disk graphs where the connection function is the indicator of a ball centred at the origin \cite{clark1991,haenggi2009}. In the following, we will be interested in the expected betweenness centrality of some node $\kappa$ found at position $\mathbf{r}_{\kappa}$ in a network formed under the above assumptions inside a disk domain $\mathcal{{D}}$.

\section{A Continuum Limit}

For the sake of mathematical tractability and in order to approximate a dense network, we consider only the continuum limit $\rho\rightarrow\infty$, where the connection range vanishes (which is realistic in the dense regime) such that $\beta\to\infty$; this scenario mimics a connected graph where all nodes on any straight line between any two points lie on the shortest path that links the two respective endpoint nodes.

We therefore seek the continuum analogue of Eq. (\ref{e:1}). Considering the probability $\frac{1}{V}d\mathbf{r_{\mathrm{i}}}$ that some node is placed at position $\mathbf{r_{\mathrm{i}}}$ in $\mathcal{{V}}$, we have the probability $\frac{1}{V^2}d\mathbf{r_{\mathrm{i}}}d\mathbf{r_{\mathrm{j}}} \chi_{ij}(\kappa)$ that (any) node pair will simultaneously be placed at $\{\mathbf{r}_{i},\mathbf{r}_{j}\}$ and construct between itself a shortest path which passes through $\kappa$, since the characteristic function $\chi_{ij}(\kappa)$ equates to unity whenever
$\kappa$ lies on the path $i\rightarrow j$ (given by the straight
line segment $\mathbf{r}_{ij}$ that joins $\mathbf{r_{\mathrm{i}}}$
and $\mathbf{r_{\mathrm{j}}}$), and is otherwise zero. Summing this up over all possible $\{\mathbf{r}_{i},\mathbf{r}_{j}\}$ pair locations within the domain gives the expected betweenness centrality of $\kappa$ for a random node configuration in $\mathcal{{V}}$ as $\rho\rightarrow\infty$:
\begin{equation}\label{e:2}
g(\kappa)=\frac{1}{2V^{2}}\int_{\mathcal{{V}}}d\mathbf{r_{\mathrm{i}}}\int_{\mathcal{{V}}}d\mathbf{r_{\mathrm{j}}}\,\chi_{ij}(\kappa)
\end{equation}
where we take $\mathcal{{V}}=\mathcal{{D}}$ and thus $V=\pi R^{2}$. Note also
that due to the symmetry of $\mathcal{{D}}$, we describe the position of the
node $\kappa$ by its Euclidean distance $\epsilon$ from the disk's centre.

Now consider Fig. \ref{fig:one}, where we define the
scalar $\kappa_{\bot}$ as the distance of $\kappa$ from the straight line $\mathbf{r}_{ij}$. Defining the delta function $\delta\left(\kappa_{\perp}\left(\mathbf{r}_{\mathrm{i}},\mathbf{r}_{j}\right)\right)$,
we then suggest that
\begin{equation}\label{e:4}
\int_{\mathcal{{D}}}d\mathbf{r_{\mathrm{i}}}\int_{\mathcal{{D}}}d\mathbf{r_{\mathrm{j}}}\,\chi_{ij}=\int_{\mathcal{{D}}}d\mathbf{r_{\mathrm{i}}}\int_{\mathcal{{D}}}d\mathbf{r_{\mathrm{j}}}\,\delta\left(\kappa_{\perp}\right)
\end{equation}
The delta function will only contribute to the integral of Eq. (\ref{e:4}) when its argument $\kappa_{\perp}$ is a zero of $\delta\left(\kappa_{\perp}\right)$. As such, if we then describe $\kappa_{\perp}$ such that it has a unique zero whenever $\kappa$ lies on the path $i\rightarrow j$, integrating $\delta\left(\kappa_{\bot}\right)$ over the
space of all node pairs $\{\mathbf{r}_{i},\mathbf{r}_{j}\}$ should
return $g(\kappa)$
as required.

\subsection*{An Expression for $\kappa_{\bot}$}

Fig. 1 shows $\kappa$ located a distance $\epsilon$ from the centre of $\mathcal{{D}}$, with the coordinate system centred on $\kappa$ and orientated such that the disk centre is at $(-\epsilon,0)$. Considering nodes $i$ and $j$ at distances $r_i$ and $r_j$ from $\kappa$ respectively, we have that the internal angles $\phi_i$, $\phi_j$ and $(\theta_j - \theta_i)$ sum to $\pi$. The perpendicular distance $\kappa_\perp$ from $\kappa$ to the line $\mathbf{r}_{ij}$ then satisfies both
\begin{equation}\label{e:101}
\frac{\kappa_\perp}{r_i} = \sin(\phi_i)
\end{equation}
and
\begin{equation}\label{e:104}
\frac{\kappa_\perp}{r_j} = \sin(\phi_j)
\end{equation}
Adding the above and taking small angle approximations (since we are interested in the case where $ \kappa_\perp\ll 1$) we have that
\begin{equation}\label{e:102}
\phi_i + \phi_j= \pi - \theta_j + \theta_i =\kappa_\perp \left( \frac{1 }{r_i} + \frac{1 }{r_j} \right)
\end{equation}
whenever $\kappa_{\bot}\ll1$. This approximation presents a unique
zero of $\kappa_{\bot}$ whenever $\theta_{i}-\theta_{j}+\pi=0$, allowing\begin{eqnarray}\label{e:8}
\delta\left(\kappa_{\bot}\right) & = & \delta\left(\frac{\theta_{i}-\theta_{j}+\pi}{\frac{1}{r_{i}}+\frac{1}{r_{j}}}\right)\nonumber\\
 & = & \delta\left(\theta_{i}-\theta_{j}+\pi\right)\left(\frac{1}{r_{i}}+\frac{1}{r_{j}}\right)
\end{eqnarray}
due to the trivial scaling laws of the delta function. Eq. (\ref{e:4}), a double volume integral, becomes a quadruple integral
\begin{eqnarray}\label{e:9}
g(\epsilon) & = & \frac{1}{2V^{2}}\int_{\mathcal{{D}}}d\mathbf{r_{\mathrm{i}}}\int_{\mathcal{{D}}}d\mathbf{r_{\mathrm{j}}}\, \chi_{ij}\left( \kappa \right) \nonumber\\
 & = & \frac{1}{2V^{2}}\int_{0}^{2\pi}d\theta_{i}\int_{0}^{2\pi}d\theta_{j}\nonumber\\
 &  & \int_{0}^{r(\theta_{i})}r_{i}dr{}_{i}\int_{0}^{r(\theta_{j})}r_{j}dr{}_{j}\delta\left(\kappa_{\bot}\right)
 \end{eqnarray}
Taking $r(\theta)=\sqrt{R^{2}-\epsilon^{2}\sin^{2}(\theta)}-\epsilon\cos\left(\theta\right)$,
the polar equation of the circle bounding $\mathcal{{D}}$, we have 
\begin{eqnarray}\label{e:901}
g(\epsilon) & = & \frac{1}{2V^{2}}\int_{0}^{2\pi}d\theta_{i}\int_{0}^{2\pi}d\theta_{j}\delta\left(\theta_{i}-\theta_{j}+\pi\right)\nonumber\\
 &  & \int_{0}^{r(\theta_{i})}r_{j}dr{}_{j}\int_{0}^{r(\theta_{j})}\left(\frac{1}{r_{i}}+\frac{1}{r_{j}}\right)r_{i}dr{}_{i}\nonumber\\
& = & \frac{1}{2V^{2}}\int_{0}^{2\pi}d\theta_{i}\int_{0}^{2\pi}d\theta_{j}\delta\left(\theta_{i}-\theta_{j}+\pi\right)\nonumber\\
 &  & \left(r(\theta_{i})\frac{r^{2}(\theta_{j})}{2}+r(\theta_{j})\frac{r^{2}(\theta_{i})}{2}\right)
 \end{eqnarray}
Integrating the delta function, we have
 \begin{eqnarray*}\label{e:902}
g(\epsilon) & = & \frac{1}{4V^{2}}\int_{0}^{2\pi}d\theta_{i}r(\theta_{i})r(\theta_{i}+\pi)\left(r(\theta_{i})+r(\theta_{i}+\pi)\right)\nonumber\\
 & = & \frac{1}{2V^{2}}\int_{0}^{2\pi}d\theta_{i}\left(R^{2}-\epsilon^{2}\right)\sqrt{R^{2}-\epsilon^{2}\sin^{2}\left(\theta_{i}\right)}\end{eqnarray*}
leaving
\begin{equation}\label{e:10}
g(\epsilon)=\frac{2\left(R^{2}-\epsilon^{2}\right)}{\pi^{2}R^{3}}E\left(\frac{\epsilon}{R}\right)
\end{equation}
where
\begin{equation}\label{e:11}
E\left(k\right)=\int_{0}^{\pi/2}d\theta\sqrt{1-k^{2}\sin^{2}\left(\theta\right)}
\end{equation}
is the complete elliptic integral of the second kind (which is related to the perimeter of an ellipse \cite{adlaj2012}). We normalise this to $g^{\star}(\epsilon)$ by dividing Eq. (\ref{e:10}) by its maximum value (such that $g^{\star}(\epsilon)g\left(0\right)=g\left(\epsilon\right)$) to obtain our main result 
\begin{eqnarray}
g^{\star}(\epsilon) =\frac{2}{\pi}\left(1-\epsilon^{2}\right)E\left(\epsilon\right)
\label{e:12}
\end{eqnarray}
with $\epsilon$ in units of $R$ (and with the betweenness now an element of the unit interval).

Elliptic integrals cannot be swiftly visualised, so for clarification we can expand Eq. (\ref{e:12}) near the origin (i.e. when $\epsilon \ll 1$) to obtain
\begin{equation}
g^\star (\epsilon\ll1) = 1- \frac{5 \epsilon^2}{R^2} + \frac{13 \epsilon^4}{64 R^4} + \mathcal{O}(\epsilon^6)
\end{equation}
while near the boundary (i.e. when $\epsilon \approx R$)
\begin{equation}
g^\star (\epsilon\approx R) = \frac{4(R-\epsilon)}{\pi R} + \mathcal{O}((R-\epsilon)^2)
\end{equation}
which implies a quadratic scaling of betweenness near the centre, and a linear scaling near the periphery.

\begin{figure}
\noindent \begin{centering}
\includegraphics[scale=0.325]{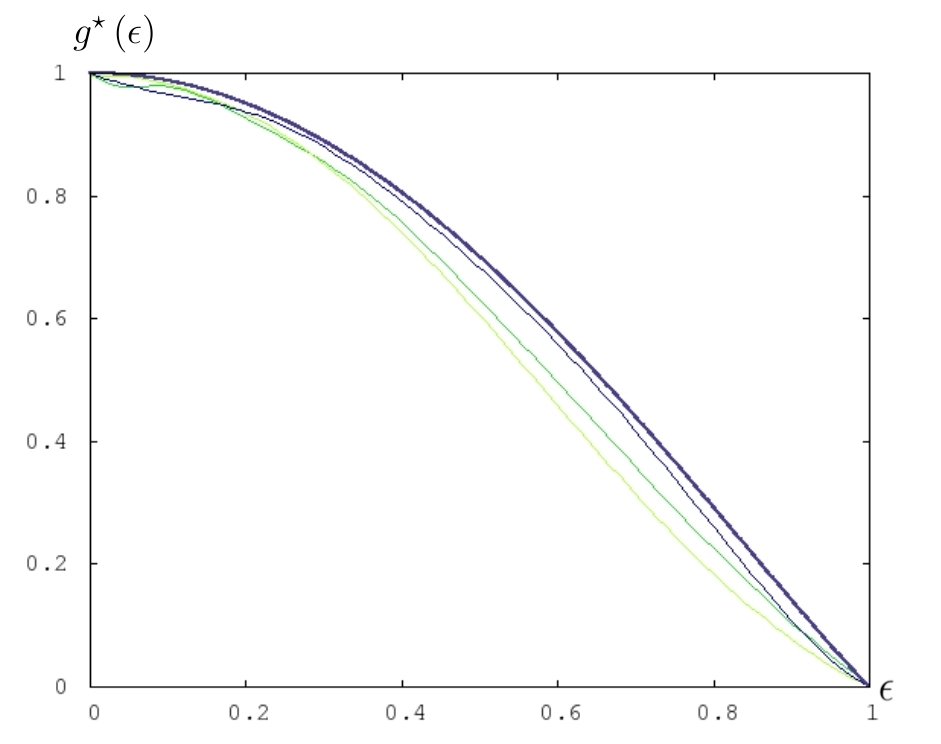}
\par\end{centering}
\caption{(Colour Online) Monte Carlo simulations: A plot of the normalised expected betweenness centrality of a node in $\mathcal{{D}}$ as a function of its distance $\epsilon$ from the centre for $\rho=10,50$ and $500$ (bottom to second top curves respectively) with Eq. (\ref{e:12}) the thicker line at the top (taking $R=1$). The finite density curves approach the limit $g^{\star}$ as $\rho\rightarrow\infty$. We sample $5000$ random graphs.}
\label{fig:three}
\end{figure}

\section{Monte Carlo Simulations} Fig. \ref{fig:two} (top left graphic) shows that the betweenness $\gamma(\kappa)$ of nodes situated in the bulk of the disk is typically high. Binning the centrality in small increments of displacement from the centre of $\mathcal{{D}}$ and averaging over many network realizations, we can plot the expectation $\mathbb{E}(\gamma\left(\epsilon\right))$, and the result is shown in Fig. \ref{fig:three}, demonstrating how $\mathbb{E}(\gamma)$ at finite densities approaches our continuum approximation. In these simulations we take $\beta$ to be the largest value required for full network connectivity \cite{orestis2013,cef2012,dettmann2007,orestis2014}, and increase $\rho$ from $10$ to $500$, each time evaluating the betweenness using Brandes' algorithm \cite{brandes2001}.

We observe that the continuum prediction is slowly reached by our numerical simulations, with only small discrepancies.
Quantifying the rate of convergence as well as the nature of these discrepancies is beyond the scope of this paper and is deferred to future work.

\section{Discussion}
By estimating betweenness based on domain location using Eq. (\ref{e:12}), nodes avoid the costly operation of repeated centrality computation throughout the network's battery-limited lifetime. At the moment, single or two-hop neighbourhood information is used in place of betweenness metrics, entirely due to the impracticality of its computation \cite{brandes2001}. This allows a range of novel, sophisticated features to be employed in future dense \textit{ad hoc} networks, which we now discuss in more detail.

\subsection{Cluster Head Node Election}

In order to minimise energy consumption, \textit{ad hoc} networks commonly group nodes into local clusters (usually defined by their inter-cluster hop distance) and elect a `cluster head node' for each partition \cite{liu2012}. The cluster head node (CH) then transmits to the distant base station (BS) on behalf of its cluster, which reportedly reduces total energy consumption by (up to) a factor of $8$ \cite{heinzelman2000}.

The betweenness measure has been used for these purposes \cite{gupta2005}, and a number of cluster routing protocols are usually implemented. For example, the basic LEACH  (Low Energy Adaptive Clustering Hierarchy \cite{heinzelman2000}) protocol uses a random selection of cluster heads at each `round' or time-step, the nodes each taking turns in bearing the burden of cloud-access (or backhaul gateway) status, or, alternatively, EECS (Energy Efficient Clustering Scheme \cite{liu2012}), which requires nodes to broadcast their remaining power to their first-degree neighbours, asking machines that find themselves with the most battery power amongst their one-hop partners to then elect themselves to CH status.

However, in large networks using a vanishing transmitter range these protocols don't work: far too many cluster heads get elected due to the huge node numbers and the efficiency problem that this technique is trying to mitigate re-arises. Potentially increasing transmitter range could resolve the problem (since the usual techniques are based on one-hop access to the head node), though this introduces interference problems, forcing the search for another solution.

Betweenness is a possible alternative election criteria (where the network is considered a single connected cluster) since it is proportional to power consumption (due to the expected increase in routing load, unlike most other centrality measures), allowing idle boundary nodes to act as cluster heads whenever power minimisation is preferred, or busy domain-center nodes whenever optimisation of node-to-node communication overheads is tasked. Knowledge of betweenness as a function of position helps in the selection of positions which, when occupied by nodes, results in CH election. In static networks this requires increasing battery resources for these stations; in mobile networks this allows nodes to use their position to trigger BS contact (perhaps at for $\epsilon\ll1$), perhaps using GPS facilities or even through measuring there current routing load.

Note also that, based on the intuition "central nodes are easier to reach", communication-based resource consumption is minimised whenever high-betweenness nodes are, in general, used as cluster heads.
\subsection{Boundary Detection}
Eq. (\ref{e:12}) gives a surface whose minimum points indicate corners, edges (and potentially faces) of the domain (see Fig. \ref{fig:two}). Boundary detection is an important field in \textit{ad hoc} network engineering, with various applications \cite{wang2006,dong2009,chen2012}. One potential use of betweenness as a boundary detector is for mitigation of the so called boundary effect phenomenon \cite{cef2012}, where high-density network connectivity is hampered through nodes becoming isolated near the domain peripheries due to a loss of the usually available full solid angle for transmission in the relevant domain dimension.
One potential mitigation technique is to increase the node transmit power at the domain boundary: by potentially using a typical node-to-node connection range $r_{0}$
\begin{equation}
r_{0}=\frac{1}{\sqrt{\beta}}
\end{equation}
where $\beta$ is a function of $\epsilon$
\begin{equation}
\beta\left(\epsilon\right)=f\left(g^{\star}\left(\epsilon\right)\right)
\end{equation}
we can harness some spare power in the relatively idle boundary nodes (detected using Eq. (\ref{e:12})), increasing machine transmit power appropriately with betweenness. This does not require the sharing of routing tables or other connectivity information, since betweenness is directly proportional to the node's current routing tasks. Finding the optimal function of the betweenness (or perhaps other centrality measures) is beyond the scope of this paper, and we defer its treatment to a later study.

\section{Conclusion}
As wireless devices and sensors become smarter, statistical methods involving low communication overheads are increasingly being developed and implemented to improve network performance. In this paper we have revisited the graph theoretic concept of betweenness centrality, a measure of how many shortest paths run through a given node, and have evaluated it in closed form in an infinitely dense random geometric network bounded inside a disk. Of course, nodes near the centre of the domain typically have more shortest paths running though them, and hence display a higher betweenness centrality, while nodes near the edge of the domain are typically used less and hence have a lower betweenness.
The quantitative formula (\ref{e:12}) presented herein, however, not only described in detail this behaviour but can also be used \textit{inter alia} for cluster head election and boundary detection within a network well modelled by a random graph (such as an ad hoc network, as discussed in the text).
The above motivates further investigations into the use of betweenness centrality in smart wireless communications under relaxed limits e.g. finite density and/or other connection models e.g. the unit disk scenario. Significantly, we next intend to focus on understanding features unique in non-convex domains, illustrated (for example) in the bottom right panel of Fig. \ref{fig:two}, where shortest paths typically route round central obstacles: this would constitute a move toward a complete analytic understanding of betweenness centrality in random geometric networks, of importance to the engineering and mathematics communities alike.

\section*{Acknowledgements}
The authors would like to thank the directors of the Toshiba Telecommunications Research Laboratory and the Centre for Doctoral Training in Communications Engineering at the University of Bristol for their continued support.

\end{document}